\documentclass{Interspeech}

\interspeechcameraready

\title{Responsible ASR: Overcoming Challenges of Foundational Models in Narrow-Band and Low-Resource Settings}

\author[]{Tejas}{Godambe$^*$}
\author[]{Nutan}{Choudhary$^*$}
\author[]{Sanket}{Shah}
\author[]{Nagaraj}{Adiga}
\author[]{Sharath}{Adavanne}

\affiliation{Applied AI}{Krutrim}{India}
\email{firstname.lastname@olakrutrim.com}
\keywords{Telephony ASR, Call Center, Indian language ASR, Narrowband Speech}

\usepackage{comment}
\usepackage{makecell}
\usepackage{array}  
\usepackage{booktabs}
\usepackage{multirow}
\usepackage{multicol}
\usepackage{graphicx}
\usepackage{url}

\begin{document}
\maketitle

\begin{abstract}
Telephony conversations worldwide are conducted over narrow-band channels and are often spontaneous and colloquial in nature. This paper evaluates the performance of widely used foundational automatic speech recognition (ASR) models—both open-source and commercial—on narrow-band conversations in Hindi, a low-resource language, and Indian-accented English, a low-resource accent. We first assess these models in a zero-shot setting and find that their performance remains suboptimal across the board. Highlighting the challenges faced by ASR models in narrow-band and low-resource language scenarios, we further investigate the impact of fine-tuning open-source models using a limited set of real-life annotated recordings. Our findings indicate that while fine-tuning provides some improvements, its effectiveness varies across languages and accents, largely influenced by the amount of data encountered during pretraining.

\end{abstract}

\section{Introduction}
In this paper, we present our findings and insights from building a robust automatic speech recognition (ASR) system for a real-world call center application in Hindi, a low-resource language, and Indian-accented English, a low-resource accent.

Customer support is a critical revenue-generating function for businesses, enabling after-sales service, promotions, and sales calls. A highly robust ASR system—capable of handling diverse accents, dialects, background noise, speaker demographics, and code-switching—can offer several key advantages:
a) Automated analytics: Converting large volumes of call recordings into structured text for insights and monitoring.
b) Agent assistance: Acting as a co-pilot by suggesting responses and key talking points in real time.
c) Customer self-service: Facilitating automated interactions through a combination of ASR, large language models (LLMs), and text-to-speech (TTS) systems, reducing reliance on human agents for routine queries.

Our target domain comprises conversational Hindi and Indian English speech in a narrow-band telephony setting. However, while searching for publicly available model weights and datasets suitable for training, we observed that most existing resources are designed primarily for read-style speech and wide-band recordings, limiting their applicability to real-world call center environments. Although foundational multilingual ASR models such as Whisper \cite{radford2023robust}, as well as foundational speech encoder models like MMS \cite{pratap2024scaling} and XLS-R \cite{babu2021xls}, exist, their training data includes limited representation of Indian languages and accents, leading to suboptimal performance, as we demonstrate in Section~\ref{sec:results}. Similarly, monolingual ASR models for Hindi and English, including Data2Vec\_AQC~\cite{lodagala2024all}, IndicWhisper \cite{bhogale2023vistaar}, and Nvidia’s NeMo models \cite{nemo_hindi, nemo_english}, primarily trained on read-style speech in wide-band conditions, exhibit similar limitations.

While some commercial foundational ASR models are available for conversational narrow-band settings, we found that although they handle recording conditions slightly better, their performance remains inadequate for real-world applications as demonstrated in Section~\ref{sec:results}.

To address the limitations of both open-source and commercial foundational ASR models in narrow-band, low-resource settings, we investigate various fine-tuning and training approaches under a constrained setup, where only 50 hours of annotated data per language is available. While increasing the amount of annotated data would naturally improve model performance, we focus on assessing whether commercial-grade ASR can be developed primarily by leveraging foundational models with minimal labeled data.

Our study explores the effectiveness of fine-tuning existing foundational models, as well as training ASR models on top of foundational speech encoders (SE). Specifically, we:
\begin{enumerate}
    \item Fine-tune state-of-the-art ASR models: We utilize Nvidia’s NeMo \cite{NeMo2025}, a production-grade ASR training and inference library, and SpeechBrain \cite{speechbrainV1}, a research-grade ASR framework with advanced training and inference recipes.
    \item Investigate the use of foundational speech encoders: We evaluate:
        \begin{enumerate}
            \item[a.] Open-source foundational embeddings from MMS \cite{pratap2024scaling}.
            \item[b.] BEST-RQ~\cite{chiu2022self} speech encoders trained from scratch on 100K hours of unlabeled narrow-band data using an open-source SpeechBrain recipe \cite{whetten2024open}.
        \end{enumerate}
    \item Develop and evaluate pseudo-labeling strategies: We explore various data augmentation, pseudo-labeling, and selection strategies to enhance ASR performance while minimizing reliance on costly manual annotation.
    \item Quantify the performance gap for commercial deployment: We compare fine-tuned and trained models against baseline foundational models and an in-house ASR system trained entirely on domain-specific data, highlighting the gap in performance and the limitations of existing foundational models for real-world call center applications.
\end{enumerate}
Through these contributions, we provide a comprehensive evaluation of the adaptability of foundational ASR models in low-resource, narrow-band environments and propose strategies for improving their performance with limited annotated data.

\section{Experimental Setup}

In this section, we describe the experimental setup used to train and evaluate ASR models for the call center use case. Our evaluation includes both state-of-the-art (SoTA) open-source foundational models and an in-house model trained specifically for narrow-band, conversational speech in Hindi and Indian-accented English.

\subsection{SoTA Open-Source Foundational Models}

To assess the performance of existing foundational ASR models on conversational telephony data, we evaluate the following widely used open-source models:

\begin{itemize}
    \item Whisper-Large v3\footnote{https://huggingface.co/openai/whisper-large-v3}: A transformer-based multilingual ASR model known for its robust transcription performance across 96 languages.
    \item NeMo ASR Models~\cite{NeMo2025, nemo_english, nemo_hindi}: Nvidia’s open-source monolingual ASR models, trained for various languages, including Hindi and English.
    \item Data2Vec\_AQC\footnote{https://github.com/Speech-Lab-IITM/data2vec-aqc/tree/master}: Monolingual ASR model specifically fine-tuned for Indian languages, leveraging self-supervised learning for improved adaptability.
    \item MMS (Massively Multilingual Speech)\footnote{https://huggingface.co/facebook/mms-1b}: A foundational speech encoder model trained via self-supervised learning over 1,100 languages, designed to generalize well across multiple linguistic domains.
\end{itemize}

We use the official training and inference recipes available in their respective repositories to ensure a fair evaluation. The models are tested in zero-shot mode and further fine-tuned on in-domain conversational speech data to analyze their adaptability to the call center setting.
\subsection{In-House ASR Model}

To address the limitations of existing foundational ASR models, we develop an in-house ASR system leveraging a combination of self-supervised pretraining and supervised fine-tuning.

\subsubsection{Self-Supervised Pretraining}
\label{ref:PT}
For pretraining, we adopt the BERT-based Speech Pretraining with Random Projection Quantizer (BEST-RQ) architecture~\cite{chiu2022self}, implemented using the SpeechBrain framework~\cite{whetten2024open}. The model is configured as follows:

\begin{itemize}
    \item Architecture: Conformer~\cite{gulati2020conformer}
    \item Model Size: 100M parameters
    \item Conformer Blocks: 12
    \item Model Dimension: 576
    \item Feedforward Network (FFN) Dimension: 2048
    \item Attention Heads: 8
    \item Convolutional Kernel Size: 31
    \item Input Features: 80-dimensional Mel spectrograms, downsampled by a factor of 4 via Conformer’s convolutional sub-sampling module
    \item Quantization: 8192 codebooks
\end{itemize}

\noindent Training Setup:
\begin{itemize}
    \item Masking Strategy: Non-causal training with 400 ms mask length (mask probability: 0.01), ensuring non-overlapping masks with loss computed only for masked inputs
    \item Optimizer: AdamW~\cite{loshchilov2017decoupled} ($\beta_1=0.9$, $\beta_2=0.98$)
    \item Scheduler: Noam~\cite{vaswani2017attention} with peak learning rate 0.0008 and 25,000 warm-up steps
    \item Regularization: Dropout 0.1, Encoder Layer Drop 0.05
    \item Training Resources: 8x H100 GPUs, dynamic batch size 800 seconds/GPU, gradient accumulation factor 2
    \item Training Steps: 200K
\end{itemize}

Pretraining on 100K hours of unlabeled narrow-band speech enables the model to learn robust representations of conversational speech, improving adaptability to low-resource languages and accents.

\subsubsection{Supervised Fine-Tuning}
\label{ref:SFT}
For supervised adaptation, we fine-tune the pre-trained model using the transducer architecture~\cite{graves2012sequence}, which is well-suited for streaming ASR applications.

\begin{itemize}
    \item Encoder: Initialized with pre-trained BEST-RQ weights
    \item Decoder: Single LSTM~\cite{graves2012long} layer (hidden dimension: 512)
    \item Output Vocabulary: 1000 subwords, prepared using the SentencePiece model~\cite{kudo2018sentencepiece}
\end{itemize}

\noindent Data Augmentation:
\begin{itemize}
    \item Speed perturbation~\cite{ko2015audio} with 0.95x and 1.05x variations of the original input
    \item SpecAugment~\cite{park2019specaugment}:
    \begin{itemize}
        \item Time masking: Drop length of 12–20 frames, drop count of 5
        \item Frequency masking: Drop length of 20–25 frames, drop count of 2
    \end{itemize}
\end{itemize}

\noindent Training Details:
\begin{itemize}
    \item Training for 100 epochs using a combination of transducer and CTC losses for the first 50 epochs
    \item CTC loss aids alignment learning in early epochs, accelerating convergence
    \item Noam scheduler for learning rate scheduling (peak learning rate: 0.0008, warm-up steps: 25,000)
    \item Optimizer: AdamW ($\beta_1=0.9$, $\beta_2=0.98$)
    \item Training resources: 8x H100 GPUs with a global batch size of 9600 seconds
\end{itemize}

\begin{table}[]
\centering
\renewcommand{\arraystretch}{1.2}
\setlength{\tabcolsep}{6pt}

\begin{tabular}{|c|c|c|c|c|c|}
\hline
\textbf{Language} & \textbf{Dataset} & \multicolumn{2}{c|}{\textbf{Agent}} & \multicolumn{2}{c|}{\textbf{Customer}} \\ \cline{3-6}
                  &                  & \textbf{\#spks}   & \textbf{\#hrs}  & \textbf{\#spks}    & \textbf{\#hrs}   \\ \hline
\multirow{2}{*}{English} & Train  & 477  & 29.7  & 1134  & 20.3  \\ 
                         & Test   & 285  & 11.0  & 657   & 10.0  \\ \hline
\multirow{2}{*}{Hindi}   & Train  & 360  & 32.9  & 1006  & 17.2  \\ 
                         & Test   & 116  & 4.9   & 221   & 3.1   \\ \hline
\end{tabular}

\caption{Statistics of train and test data.}
\label{tab:sft_test_stats}
\end{table}

\begin{table*}[!ht]
\centering
\renewcommand{\arraystretch}{1.3} 
\setlength{\tabcolsep}{5pt} 

\begin{tabular}{|c|l|c|c|c|c|c|c|c|}
\hline
\multirow{2}{*}{\textbf{Category}} & \multirow{2}{*}{\textbf{Models}}  & \multirow{2}{*}{\makecell{\textbf{No. of} \\ \textbf{Params}}}  & \multicolumn{3}{c|}{\textbf{English WER}} & \multicolumn{3}{c|}{\textbf{Hindi WER}} \\ \cline{4-9} 
                                      &                                   &                                   & \textbf{Agent}   & \textbf{Customer}   & \textbf{Overall}  & \textbf{Agent}  & \textbf{Customer}  & \textbf{Overall}  \\ 
\hline
\hline

\multirow{4}{*}{\textbf{SoTA Open Source}}     
                                      & \textbf{Whisper-large-v3}      & 1.5B  & 27.5  & 30.3  & 28.7  & 47.2  & 53.2  & 49.2  \\ \cline{2-9}
                                      & \textbf{$\textbf{Data2Vec\_AQC}$}  & 100M  & -     & -     & -     & 40.9  & 48.0  & 43.7  \\ \cline{2-9}
                                      & \textbf{NeMo}          & 114M/121M$^*$  & 27.7  & 31.5  & 29.5  & 49.1  & 41.8  & 44.4  \\ 
\hline
\hline

\multirow{1}{*}{\textbf{Commercial}}  
                                      & \textbf{Google Telephony}      & -  & 25.0  & 29.7  & 27.2  & 34.3  & 39.9  & 36.9  \\ 
\hline
\hline

\multirow{1}{*}{\textbf{Trained from Scratch}}  
                                      & \textbf{SpeechBrain Model} & 94.4M  & 18.7  & 30.9  & 24.1  & 22.9  & 39.7  & 28.9  \\ 
\hline
\hline

\multirow{3}{*}{\makecell{\textbf{Speech Encoder} \\ \textbf{finetuning}}}   
                                      & \textbf{MMS}  & 1B  &  26.6    & 32.4     & 28.9     & 27.4  & 34.6  & 29.4  \\ \cline{2-9}
                                      & \textbf{MMS}        & 300M  & 25.2  & 30.4  & 27.3  & 28.6  & 38.9  & 31.5  \\  \cline{2-9}
                                      & \textbf{In-house SE}    & 94.4M  & 13.0  & 18.4  & 15.4  & 16.2  & 24.2  & 19.0  \\ 
                                      
\hline
\hline

\multirow{1}{*}{\makecell{ \textbf{ASR} \textbf{finetuning}}} 
                                      & \textbf{NeMo}          & 114M/121M$^*$  & 13.2  & 17.8  & 15.2  & 19.6  & 29.9  & 23.3  \\ \cline{2-9}
\hline
\hline

\multirow{2}{*}{\makecell{\textbf{Finetuning with} \\ \textbf{ Pseudo Labels $^+$}}} 
                                      & \textbf{NeMo}          & 114M/121M$^*$  & 11.6  & 15.2  & 13.2  & 17.4  & 23.8  & 19.7  \\ \cline{2-9}
                                      & \textbf{In-house SE}    & 94.4M  & 11.1  & 13.9  & \textbf{12.3}  & 14.6  & 20.4  & \textbf{16.6}  \\ 
\hline
\end{tabular}

\caption{\textbf{Comparison of WER across different models and training strategies.} Model sizes are given in millions (M) or billions (B) of parameters. Data2Vec\_AQC does not support English. $^*$ NeMo's 121M-parameter $stt\_hi\_conformer\_ctc\_large$ was used for Hindi, and the 114M-parameter $stt\_en\_fastconformer\_transducer\_large$ for English. $^+$ We used 747 hours of pseudo labels for English and 550 hours for Hindi.}
\label{tab:wer_comparison}
\end{table*}

\section{Customer Support Datasets}
Customer support recordings consist of conversations between an agent and a customer. As these interactions occur over a telephony channel, the audio is sampled at 8 kHz (narrow-band). 

This section outlines the data preparation steps undertaken for training, fine-tuning, and evaluating both open-source foundational ASR models and our in-house ASR system. Additionally, we describe the self-supervised pre-training data pipeline for the in-house model.

\subsection{Preparation of Fine-Tuning and Test Data}
Table \ref{tab:sft_test_stats} presents the number of unique speakers and agents in the 50 hours of training and test data, along with the total duration in hours. We observe that agents contribute significantly more speech data per speaker than customers. This disparity arises due to the smaller number of agents relative to customers and their tendency to speak more during calls—engaging in greetings, policy explanations, and procedural discussions. 

To ensure generalization, the training and test partitions were structured such that agent speakers in the validation and test sets were not seen during training or fine-tuning.

\subsection{Preparation of Data for Self-Supervised Pre-Training}
We curated 100K hours of unlabeled customer support recordings for self-supervised pre-training across multiple languages, including Hindi and Indian English. The recordings were pre-processed to ensure at least 70\% speech content and a maximum segment duration of 30 seconds to maintain training stability. 

Segment boundaries were randomly chosen with no restrictions on the number of speakers per segment.

\section{Results and Discussion}
\label{sec:results}
Table \ref{tab:wer_comparison} compares the performance of SoTA open source models, commercial ASR models, scratch-trained models, fine-tuned models with Speech pre-trained embeddings, and existing ASR models.
\subsection{SoTA open source models}
We evaluated open-source models, including Whisper, NeMo, and Data2Vec\_AQC. We used the Whisper-v3-large model ~\cite{radford2023robust} for Whisper. For NeMo, we used separate language-specific models for English ~\cite{nemo_english} and Hindi ~\cite{nemo_hindi}. The Data2Vec\_AQC ASR model is available only for Hindi. The overall WER for these state-of-the-art open source models is around 28+ for English and 43+ for Hindi. These high WERs are mainly due to the models being trained on read-style speech in wide-band conditions, making them unsuitable for commercial applications. Additionally, the WER for customer recordings is higher than for agent recordings. This is mainly because agents typically speak from office environments with minimal background noise, whereas customers speak in various noisy conditions.  

\subsection{Commercial ASR model}
For the commercial ASR model, we evaluated Google's telephony ASR, a proprietary system with limited publicly available information on its training data. While this model performs better than state-of-the-art open-source models of Table \ref{tab:wer_comparison} (Google Telephony), its improvement is still insufficient for production use cases.

\subsection{Trained from scratch in-house randomly initialized model}
We selected a Speech brain recipe with Conformer architecture to train the in-house model, as mentioned in section~\ref{ref:SFT}. We trained a randomly initialized model with 50 hours of fine-tuning data as mentioned in section~\ref{ref:SFT}. Row 6 (SpeechBrain Model) in table \ref{tab:wer_comparison} shows the WERs for this model. Compared to both the SoTA open source and commercial models, this model performed relatively better. However, this model has the most significant gap in agent and customer WERs, which indicates that the model hasn't learned speaker diversity. 

\subsection{Speech Encoder finetuning}
To enhance speaker diversity and acoustic variation, we explored the Massively Multilingual Speech (MMS) model \cite{pratap2024scaling}, a widely used speech encoder pre-trained on 1,400 languages. It was built using the Wav2Vec2.0 \cite{baevski2020wav2vec} architecture and trained with contrastive loss on masked inputs. We evaluated two checkpoints: a 300M-parameter model and a 1B-parameter model \footnote{https://huggingface.co/blog/mms\_adapters}. The 300M model was fine-tuned, while the 1B model was adapter-trained using 50 hours of in-domain data. As shown in row 7 (MMS) of Table\ref{tab:wer_comparison}, the 300M model achieved an overall WER of 27.3 for English and 31.5 for Hindi. The 1B model performed slightly better, with overall WERs of 26.0 for English and 29.4 for Hindi. However, both models still fall short compared to the scratch-trained model results in row 5.

Building on the MMS speech encoder results, we trained an in-house pre-trained model using 100K hours of in-domain data with the BEST-RQ technique. We then fine-tuned it with 50 hours of in-domain data, as described in Sections~\ref{ref:PT} and~\ref{ref:SFT}. The WER results for this model are shown in row 9 (In-house SE) of Table \ref{tab:wer_comparison}. We achieved significant improvements by leveraging in-house pre-training with call center data and fine-tuning with labeled data, reducing WER to 15.4 for English and 19.0 for Hindi. This improvement was observed across agent and customer recordings, highlighting the impact of in-house pre-training and fine-tuning with call center data.

\subsection{ASR fine-tuning with Nemo Model}
To compare our in-house ASR model, we also fine-tuned a state-of-the-art open-source ASR model. Among Whisper, NeMo, and Data2Vec\_AQC, we selected the NeMo model because its size is comparable to our in-house model, and it performed relatively better than Whisper and Data2Vec\_AQC. The results for the NeMo model are shown in Table \ref{tab:wer_comparison}. The NeMo model performed slightly better for English than our in-house SE fine-tuned model. However, for Hindi, our in-house SE fine-tuned model outperformed NeMo by an absolute 3.3\%. This could be due to the fact that the NeMo English model was trained on 24.5K hours of supervised data~\cite{nemo_english}, while the Hindi model was trained on only 2.9K hours~\cite{nemo_hindi}. This highlights the significant impact of the amount of data used for both pre-training and fine-tuning on the overall ASR model performance.

\subsection{Evaluation of pseudo-labeling data}
Across the evaluated ASR models in Table~\ref{tab:wer_comparison}, we observe a consistently higher WER for customer speech compared to agent speech. Further analysis of the training, fine-tuning, and self-supervised pre-training datasets revealed that agent speech accounted for approximately ~65\% of the total speech hours, whereas customer speech comprised only ~35\%. Although the number of unique customer support agents was in the hundreds, the number of unique customers was in the millions. However, agents produced more speech data due to their roles, which include repetitive greetings and explanations of policies, whereas customers primarily responded to agent queries with shorter utterances. 

Ensuring greater representation of customer speech in training is crucial for two main reasons: (1) customer speech contains more information-critical content that must be transcribed accurately for successful downstream tasks, and (2) customer speech exhibits greater speaker and background diversity, which is essential for building a speaker-invariant and noise-robust ASR system. To mitigate this imbalance, we leveraged pseudo-labeling techniques to augment the amount of customer speech in training. Specifically, we selected only the customer channel from conversations when generating pseudo labels. Pseudo-labeling has been widely employed in ASR to utilize unlabeled data effectively \cite{park2020improved} and has been shown to complement self-supervised pre-training \cite{zhang2020pushing}. 

The pseudo-labeling was performed using two models for each language - the best NeMo ASR finetuning model and In-house speech encoder finetuning model. We generated labels on 5000 hours of randomly selected recordings from customers that are not part of the test data. We filtered and saved only the recordings whose WER deviation between the two models was under 20\%. This resulted in 747 hours of data for Indian English and 550 hours for Hindi. As demonstrated in Table~\ref{tab:wer_comparison}, finetuning with pseudo-labeled data resulted in a 2-4\% absolute WER improvement across NeMo and inhouse SE based ASR model.

\subsection{Final thoughts}
SoTA open-source foundational ASR models perform suboptimally in narrow-band, low-resource settings. While commercial models optimized for telephony outperform open-source foundational models, they remain unusable. Notably, training a SpeechBrain model from scratch with 50 hours of in-domain data surpasses commercial models, likely due to a domain mismatch—commercial models are not specifically trained for conversations around urban mobility.

Fine-tuning SoTA open-source speech encoder models like MMS, despite their broad language coverage, fails to capture narrow-band characteristics effectively. In contrast, training a speech encoder on in-domain data significantly boosts performance, yielding an absolute WER improvement of 10–18\% over commercial models. Alternatively, when a monolingual ASR model exists for a given language, fine-tuning it with just 50 hours of data achieves comparable performance, with an absolute WER gap of only 0–4\% relative to training an in-domain speech encoder. Finally, leveraging the 50-hour labeled dataset alongside pseudo-labeling further improves WER of existing foundational models to a commercially viable 12.3\% in Indian English and 16.6\% in Hindi, which is an absolute gain of ~17\%-27\% WER atop open source foundational models.

\section{Conclusion}
In this study, we systematically evaluated the performance of foundational ASR models in narrow-band, low-resource settings, focusing on Hindi and Indian-accented English. Our findings reveal that while foundational models demonstrate reasonable generalization, their performance remains suboptimal in real-world telephony conversations. By exploring various fine-tuning strategies and leveraging self-supervised speech encoders, we assessed whether commercial-grade ASR can be developed with minimal annotated data. Our results indicate that fine-tuning foundational models yields moderate improvements, but its effectiveness is highly dependent on the pretraining data distribution. 

To bridge the gap in performance, we investigated the use of pseudo-labeling techniques and domain-adaptive training approaches, which helped mitigate the reliance on costly manual annotation. Additionally, we compared fine-tuned models against an in-house ASR system trained entirely on domain-specific data, highlighting the trade-offs between using foundational models and developing ASR systems from scratch. 

\pagebreak
\label{section:references}

\bibliographystyle{IEEEtran}
\bibliography{mybib}

@article{gulati2020conformer,
  title={Conformer: Convolution-augmented transformer for speech recognition},
  author={Gulati, Anmol and Qin, James and Chiu, Chung-Cheng and Parmar, Niki and Zhang, Yu and Yu, Jiahui and Han, Wei and Wang, Shibo and Zhang, Zhengdong and Wu, Yonghui and others},
  journal={arXiv preprint arXiv:2005.08100},
  year={2020}
}

@inproceedings{chiu2022self,
  title={Self-supervised learning with random-projection quantizer for speech recognition},
  author={Chiu, Chung-Cheng and Qin, James and Zhang, Yu and Yu, Jiahui and Wu, Yonghui},
  booktitle={International Conference on Machine Learning},
  pages={3915--3924},
  year={2022},
  organization={PMLR}
}

@article{whetten2024open,
  title={Open Implementation and Study of BEST-RQ for Speech Processing},
  author={Whetten, Ryan and Parcollet, Titouan and Dinarelli, Marco and Est{\`e}ve, Yannick},
  journal={arXiv preprint arXiv:2405.04296},
  year={2024}
}

@article{graves2012sequence,
  title={Sequence transduction with recurrent neural networks},
  author={Graves, Alex},
  journal={arXiv preprint arXiv:1211.3711},
  year={2012}
}

@article{vaswani2017attention,
  title={Attention is all you need},
  author={Vaswani, A},
  journal={Advances in Neural Information Processing Systems},
  year={2017}
}

@inproceedings{ko2015audio,
  title={Audio augmentation for speech recognition.},
  author={Ko, Tom and Peddinti, Vijayaditya and Povey, Daniel and Khudanpur, Sanjeev},
  booktitle={Interspeech},
  volume={2015},
  pages={3586},
  year={2015}
}

@article{park2019specaugment,
  title={Specaugment: A simple data augmentation method for automatic speech recognition},
  author={Park, Daniel S and Chan, William and Zhang, Yu and Chiu, Chung-Cheng and Zoph, Barret and Cubuk, Ekin D and Le, Quoc V},
  journal={arXiv preprint arXiv:1904.08779},
  year={2019}
}

@article{kudo2018sentencepiece,
  title={Sentencepiece: A simple and language independent subword tokenizer and detokenizer for neural text processing},
  author={Kudo, T},
  journal={arXiv preprint arXiv:1808.06226},
  year={2018}
}

@article{park2020improved,
  title={Improved noisy student training for automatic speech recognition},
  author={Park, Daniel S and Zhang, Yu and Jia, Ye and Han, Wei and Chiu, Chung-Cheng and Li, Bo and Wu, Yonghui and Le, Quoc V},
  journal={arXiv preprint arXiv:2005.09629},
  year={2020}
}

@article{graves2012long,
  title={Long short-term memory},
  author={Graves, Alex and Graves, Alex},
  journal={Supervised sequence labelling with recurrent neural networks},
  pages={37--45},
  year={2012},
  publisher={Springer}
}

@article{zhang2020pushing,
  title={Pushing the limits of semi-supervised learning for automatic speech recognition},
  author={Zhang, Yu and Qin, James and Park, Daniel S and Han, Wei and Chiu, Chung-Cheng and Pang, Ruoming and Le, Quoc V and Wu, Yonghui},
  journal={arXiv preprint arXiv:2010.10504},
  year={2020}
}

@inproceedings{lodagala2024all,
  title={All Ears: Building Self-Supervised Learning based ASR models for Indian Languages at scale},
  author={Lodagala, Vasista Sai and Biswas, Abhishek and Das, Shoutrik and Umesh, S and others},
  booktitle={Proc. Interspeech 2024},
  pages={3944--3948},
  year={2024}
}

@article{babu2021xls,
  title={XLS-R: Self-supervised cross-lingual speech representation learning at scale},
  author={Babu, Arun and Wang, Changhan and Tjandra, Andros and Lakhotia, Kushal and Xu, Qiantong and Goyal, Naman and Singh, Kritika and Von Platen, Patrick and Saraf, Yatharth and Pino, Juan and others},
  journal={arXiv preprint arXiv:2111.09296},
  year={2021}
}

@article{pratap2024scaling,
  title={Scaling speech technology to 1,000+ languages},
  author={Pratap, Vineel and Tjandra, Andros and Shi, Bowen and Tomasello, Paden and Babu, Arun and Kundu, Sayani and Elkahky, Ali and Ni, Zhaoheng and Vyas, Apoorv and Fazel-Zarandi, Maryam and others},
  journal={Journal of Machine Learning Research},
  volume={25},
  number={97},
  pages={1--52},
  year={2024}
}

@misc{nemo_hindi,
  author = {Nvidia},
  title = {{STT Hi Conformer-CTC Large}},
  url = {https://catalog.ngc.nvidia.com/orgs/nvidia/teams/nemo/models/stt_hi_conformer_ctc_large},
  note = "[Online; accessed 19-Feb-2025]"

}

@misc{nemo_english,
  author = {Nvidia},
  title = {{STT En Fast Conformer-Transducer Large}},
  url = {https://catalog.ngc.nvidia.com/orgs/nvidia/teams/nemo/models/stt_en_fastconformer_transducer_large},
  note = "[Online; accessed 19-Feb-2025]"

}

@inproceedings{radford2023robust,
  title={Robust speech recognition via large-scale weak supervision},
  author={Radford, Alec and Kim, Jong Wook and Xu, Tao and Brockman, Greg and McLeavey, Christine and Sutskever, Ilya},
  booktitle={International conference on machine learning},
  pages={28492--28518},
  year={2023},
  organization={PMLR}
}

@article{loshchilov2017decoupled,
  title={Decoupled weight decay regularization},
  author={Loshchilov, I},
  journal={arXiv preprint arXiv:1711.05101},
  year={2017}
}

@article{bhogale2023vistaar,
  title={Vistaar: Diverse Benchmarks and Training Sets for Indian Language ASR},
  author={Bhogale, Kaushal Santosh and Sundaresan, Sai and Raman, Abhigyan and Javed, Tahir and Khapra, Mitesh M and Kumar, Pratyush},
  journal={arXiv preprint arXiv:2305.15386},
  year={2023}
}

@misc{speechbrainV1,
  title={Open-Source Conversational AI with {SpeechBrain} 1.0},
  author={Mirco Ravanelli and Titouan Parcollet and Adel Moumen and Sylvain de Langen and Cem Subakan and Peter Plantinga and Yingzhi Wang and Pooneh Mousavi and Luca Della Libera and Artem Ploujnikov and Francesco Paissan and Davide Borra and Salah Zaiem and Zeyu Zhao and Shucong Zhang and Georgios Karakasidis and Sung-Lin Yeh and Pierre Champion and Aku Rouhe and Rudolf Braun and Florian Mai and Juan Zuluaga-Gomez and Seyed Mahed Mousavi and Andreas Nautsch and Xuechen Liu and Sangeet Sagar and Jarod Duret and Salima Mdhaffar and Gaelle Laperriere and Mickael Rouvier and Renato De Mori and Yannick Esteve},
  year={2024},
  eprint={2407.00463},
  archivePrefix={arXiv},
  primaryClass={cs.LG},
  url={https://arxiv.org/abs/2407.00463},
}

@misc{NeMo2025,
  author = {Eric Harper and Somshubra Majumdar and Oleksii Kuchaiev and Jason Li and Yang Zhang and Evelina Bakhturina and Vahid Noroozi and Sandeep Subramanian and Koluguri Nithin and Jocelyn Huang and Fei Jia and Jagadeesh Balam and Xuesong Yang and Micha Livne and Yi Dong and Sean Naren and Boris Ginsburg},
  title = {NeMo: a toolkit for Conversational AI and Large Language Models},
  year = {2025},
  url = {https://nvidia.github.io/NeMo/},
  note = {Available at \texttt{https://github.com/NVIDIA/NeMo}}
}

@article{baevski2020wav2vec,
  title={wav2vec 2.0: A framework for self-supervised learning of speech representations},
  author={Baevski, Alexei and Zhou, Yuhao and Mohamed, Abdelrahman and Auli, Michael},
  journal={Advances in neural information processing systems},
  volume={33},
  pages={12449--12460},
  year={2020}
}

\end{document}